# NOTA SOBRE ALGUMAS INTERPRETAÇÕES DA TEORIA DE TRIBUTAÇÃO ÓTIMA

*José Ricardo Bezerra Nogueira*[1]
*Departamento de Economia*
*Universidade Federal de Pernambuco*
*(setembro de 2021)*

**Resumo**
Esta nota discute alguns aspectos de interpretações da teoria da tributação ótima apresentadas em trabalhos recentes sobre o sistema tributário brasileiro.

**Abstract**
This note discusses some aspects of interpretations of the theory of optimal taxation presented in recent works on the Brazilian tax system.

## 1. Introdução

Recentemente, vários trabalhos discutindo o impacto distributivo do sistema brasileiro de tributos e benefícios sociais recorrem a uma interpretação crítica da teoria da tributação ótima. Entre esses trabalhos, podem-se mencionar Vianna e Silveira (2020) e Silveira, Passos, Silva e Palomo (2020), Orair e Gobetti (2019), Gobetti (2019), Avila, Martins e Conceição (2019), Orair e Gobetti (2018), Silveira, Passos e Guedes (2018), Gobetti e Orair (2017) e Gobetti e Orair (2016).

Essa interpretação, além de ser, ao nosso ver, inapropriadamente utilizada como base para criticar análises divergentes sobre o papel do sistema tributário e dos gastos sociais, está claramente equivocada em relação ao entendimento da própria teoria da tributação ótima. Nesta breve nota nos concentramos nesse último aspecto, ou seja, em realçar o entendimento errôneo da teoria conforme exposto nos referidos trabalhos.

Criticar uma teoria, e, portanto, a teoria da tributação ótima, é perfeitamente legítimo. Críticas constituem elementos importantes para o desenvolvimento de uma teoria. A própria teoria da tributação ótima, como observado pelos autores acima referidos, tem sido objeto de críticas, autocríticas e extensões ao longo do tempo, processo esse que tem possibilitado gerar um fluxo de novos *insights* a partir dos modelos básicos.

---

[1] E-mail: jrbnogueira@yahoo.com.br

As críticas, entretanto, devem refletir uma compreensão acurada da teoria considerada. Portanto, uma crítica da abordagem de tributação ótima deve necessariamente refletir um adequado entendimento do que ela efetivamente é. Nosso propósito aqui é argumentar que isso não é encontrado nas referências acima aludidas.

## 2. O argumento

O ponto central dos trabalhos acima citados é de que, na discussão do impacto distributivo da política fiscal, ênfase é dada mais ao lado dos gastos do que dos tributos:

> Ao longo das últimas quatro décadas prevaleceu na academia e no debate público em geral a ideia de que a intervenção com fins redistributivos do Estado deveria se dar pela via do gasto, mediante políticas sociais compensatórias, preferencialmente "focalizadas". (Vianna e Silveira, 2020, p.2)

> O enfrentamento da desigualdade via política fiscal pode ser empreendido tanto via tributação como gastos, mas a história revela que há uma maior aceitação do segundo do que do primeiro. (Silveira, Passos, Silva e Palomo, 2020, p.2)

Essa ênfase teria por base a ideia da existência de um *trade-off* entre eficiência e equidade, ideia essa fundamentada na teoria da tributação ótima:

> Como suporte a essa ideia, a suposta existência de um *tradeoff* insuperável entre eficiência e equidade, que deveria orientar as escolhas dos formuladores e gestores de políticas públicas [...] O fundamento dessa visão reside na chamada teoria da tributação ótima [...] (Vianna e Silveira, 2020, p.2).

> [M]esmo organismos multilaterais historicamente vinculados à ortodoxia (e, portanto, à abordagem da tributação ótima), como o Fundo Monetário Internacional (FMI), têm promovido uma revisão de seus posicionamentos, publicando relatórios e textos para discussão [...] que enfatizam efeitos negativos da desigualdade para a macroeconomia e para o crescimento econômico de longo prazo – isto é, contradizendo o *tradeoff* entre equidade e eficiência e, principalmente, defendendo abertamente a adoção de políticas tributárias progressivas para enfrentar de maneira efetiva a questão da desigualdade. (Vianna e Silveira, 2020, p.2)

A partir dessa base, a teoria da tributação ótima preconizaria que o sistema tributário fosse neutro em termos distributivos, de forma a não exacerbar os efeitos negativos do *trade-off* em relação à eficiência. Por conseguinte, essa prescrição resultaria, em última instância, em uma ênfase na utilização do lado do gasto na política fiscal como instrumento superior para lidar com a questão da equidade:

> Por influência de leituras restritivas da literatura de tributação ótima [...] construiu-se uma espécie de consenso no *mainstream* e entre *policymakers* de que a política tributária, para não introduzir distorções no sistema econômico, deveria se abster dos

> objetivos distributivos, transferindo-se ao gasto público essa função clássica da política fiscal (Gobetti e Orair, 2016, p.8).

> [S]egundo a visão dominante nos anos 90, a política tributária deveria se eximir de objetivos distributivos porque isso seria ineficiente do ponto de vista econômico. O gasto público, por outro lado, seria o instrumento adequado por meio do qual o governo poderia interferir na distribuição de renda, principalmente se bem focalizados, ou seja, direcionados aos realmente pobres. Assim, a autoridade fiscal deveria se preocupar em ampliar a neutralidade do sistema tributário, garantindo as condições mediante as quais a economia proporcionaria o máximo de arrecadação para atender aos objetivos alocativos e distributivos do governo (Gobetti, 2019, p.3).

Segundo essa interpretação, a teoria da tributação ótima ofereceria, portanto, suporte a uma prescrição de política tributária que excluiria (ou drasticamente restringiria) objetivos distributivos.

A interpretação prossegue afirmando que essa prescrição de uma política tributária "neutra" estaria ligada a uma primeira fase no desenvolvimento da teoria da tributação ótima. Desenvolvimentos recentes, a partir dos anos 1990s teriam, então, levado a uma refutação de resultados da primeira fase dessa abordagem e resgatado um papel ativo para a política tributária no enfrentamento das desigualdades.

> Três décadas se passaram desde que esses mandamentos foram estabelecidos, e tanto a concentração de renda aumentou significativamente na maior parte do mundo, quanto a reflexão acadêmica avançou, produzindo uma reavaliação das teorias e práticas em termos de tributação. Alguns novos desenvolvimentos da teoria da tributação ótima, por exemplo, têm questionado a espinha dorsal dos modelos e teoremas que se popularizaram no *mainstream* acadêmico e que deram sustentação às proposições de menor progressividade tributária e menor tributação das rendas do capital.
>
> Tanto economistas da nova geração, como Piketty e Emmanuel Saez, quanto da velha guarda, como Joseph Stiglitz, Anthony Atkinson e Peter Diamond, têm se dedicado a demonstrar, por meio de um arcabouço estritamente neoclássico, mas baseado em hipóteses e pressupostos mais realistas, que uma política tributária ótima (no sentido de maximizar o bem-estar social) pode passar por um desenho onde não só haja espaço para a progressividade tributária e a tributação do capital, como em doses superiores à que temos na atualidade. (Gobetti, 2019, p.2)

> [N]a segunda década do século XXI a Teoria da Tributação Ótima passou por uma revisão. Os autores voltaram atrás em suas recomendações e passaram a sugerir o retorno da tributação sobre os rendimentos do capital, além da continuidade de políticas de Estado via serviços públicos para reduzir as desigualdades de renda". (Avila, Martins e Conceição, 2019, p.2)

## 3. A crítica da teoria da tributação ótima no argumento

O ponto de partida da crítica apresentada é a ideia da existência de um "suposto" *trade-off* entre eficiência e equidade na base da teoria da tributação ótima, ou seja, da existência de um conflito básico entre os objetivos de eficiência e equidade.

A aceitação desse *trade-off* resultaria, por sua vez, na aceitação de uma política tributária passiva em relação ao objetivo de equidade, de forma a evitar (ou reduzir) o impacto negativo do *trade-off* sobre eficiência. Nesse sentido, a teoria da tributação ótima prescreveria, entre outras medidas:

a) Neutralidade da política tributária.

> [A] orientação é que a tributação se assente no princípio da neutralidade, de modo a minimizar os efeitos distorcivos no comportamento dos agentes. A ideia é que ações que visem redistribuição por meio do sistema tributário, tais como progressividade na tributação das rendas oriundas do trabalho e do capital, podem ser inócuas ao desestimular o trabalho dos indivíduos mais habilitados e produtivos. Assim, a TTO conclui que a distribuição de renda via sistema tributário é contraproducente (Silveira, Passos, Silva e Palomo, 2020, p.4).

b) Alíquotas marginais de imposto de renda iguais a zero.

> As sugestões preconizavam alíquotas marginais de imposto de renda iguais a zero, isenção de imposto de renda sobre os lucros e desoneração do capital, isto é, a rejeição de políticas redistributivas através da tributação. (Avila, Martins e Conceição, 2019, p.1-2).

Ou

c) Imposto sobre a renda com alíquota linear e não tributação do capital.

> [A] teoria da tributação ótima, que, originalmente, baseada no alegado *trade-off* entre equidade e eficiência e em hipóteses muito restritivas sobre o comportamento individual e a dinâmica econômica, produzia modelos extremos, em que o IR deveria ter uma alíquota linear e as rendas do capital não deveriam ser tributadas para não distorcer incentivos econômicos. (Gobetti e Orair, 2016, p.7)

> Assim, do ponto de vista histórico, o Brasil não aderiu, de fato, à onda favorável à promoção da justiça fiscal no período de consolidação dos Estados de Bem-Estar, porém, em sentido contrário, optou pelo caminho da Tributação Ótima cuja justificava seguia dois argumentos: i) suavização da progressividade ou adoção de um imposto de renda linear, e; ii) redução ou eliminação da tributação sobre as rendas oriundas do capital. (Silveira, Passos e Guedes, 2018, p.2)

> Alguns novos desenvolvimentos da teoria da tributação ótima, por exemplo, têm questionado a espinha dorsal dos modelos e teoremas que se popularizaram no

*mainstream* acadêmico e que deram sustentação às proposições de menor progressividade tributária e menor tributação das rendas do capital. (Gobetti, 2019, p.3)

## 4. A Crítica da crítica

A interpretação da teoria da tributação ótima apresentada acima sugere que a teoria parte de pressupostos frágeis e é, pelo menos até mais recentemente, contrária à utilização do sistema tributário como instrumento de redução da desigualdade. Nesta seção, argumentamos que essa visão da teoria é equivocada e distorce tanto suas bases quanto o seu propósito maior.

O primeiro passo para a análise de uma teoria é entender o seu objetivo, e não simplesmente os seus resultados. Isso permite delinear seu campo de atuação e, portanto, seus limites.[2]

Em relação à teoria da tributação ótima, seu objetivo fundamental é fornecer as bases de uma teoria de políticas ótimas, isto é, um arcabouço teórico que permita construir ferramentas úteis para a elaboração, quantificação e análise de políticas públicas levando em consideração a interrelação entre as metas almejadas e as circunstâncias que as condicionam. Em suma, desde seu início, a teoria da tributação ótima procurou estabelecer uma generalização da formalização e implementação dos problemas de otimização individuais, agora incorporando a dimensão da escolha social ótima.

O ponto de partida dessa generalização é o conflito (*trade-off*) entre eficiência e equidade. Não se trata de um "suposto" conflito, e sim de uma situação inescapável, que está na base de toda a análise econômica, em se tratando de utilização de recursos escassos. É um princípio básico da análise econômica que toda decisão tomada incorre inevitavelmente em um custo de oportunidade.

Na teoria da tributação ótima, esse *trade-off* é explicitado, a partir do Segundo Teorema Fundamental do Bem-Estar, através da demonstração de que, com exceção de transferências do tipo *lump-sum*, todo e qualquer tributo/subsídio, necessariamente envolve perdas de eficiência. A teoria da tributação ótima, então, procura caracterizar criteriosamente a natureza essencial desse trade-off de forma a sugerir meios de balancear os critérios de equidade e eficiência. Formalmente, isso é feito através da condição de que a política adotada minimize dos custos de eficiência (perda de peso morto) associados ao objetivo redistributivo (equidade) almejado.

---

[2] Para uma revisão geral da teoria da tributação ótima, ver Boadway (2012).

Dada essa caracterização geral da abordagem da tributação ótima, passamos a considerar alguns pontos específicos da crítica a ela relacionada.

*Neutralidade da política tributária*

Não decorre do arcabouço básico da teoria da tributação ótima, ao contrário do argumentado pela crítica, que haja uma prescrição de que a política tributária seja passiva relativamente ao objetivo de equidade. Assumir esse *non sequitur* a crítica revela uma inadequada interpretação da teoria. A teoria explicitamente incorpora o critério de equidade na análise de políticas ótimas e permite que se analise como diferentes tipos de tributos e diferentes estruturas de alíquotas tributárias podem ser utilizados para alcançar o objetivo de equidade, ao mesmo tempo que mostra o custo de oportunidade (em termos de eficiência) de tal objetivo.

Mais especificamente, não faz sentido afirmar que a teoria da tributação ótima argumenta que a distribuição de renda via tributação é inócua ou contraproducente.[3] Em nenhum resultado isso é encontrado. A confusão aqui é provocada por entender, erroneamente, que o reconhecimento do inevitável custo de oportunidade gerado pela tributação seja equivalente a uma demonstração da inocuidade da redistribuição via tributação. Ao contrário, nunca é demais enfatizar, a teoria da tributação ótima incorpora o critério de equidade na análise tributária e torna possível examinar criteriosamente o conflito entre equidade e eficiência.

A única prescrição de neutralidade a partir da teoria da tributação ótima (e, de forma mais geral, da teoria econômica da tributação) refere-se ao argumento de que o sistema tributário deveria ser neutro em relação aos agentes tributados, não tributando diferenciadamente contribuintes e atividades em condições similares. Esse argumento tem base na ideia de justiça tributária, incorporada na noção de equidade horizontal.

*Alíquotas marginais de imposto de renda iguais a zero*

Outra inferência equivocada da teoria da tributação ótima é afirmar que ela prescreve alíquotas marginais de imposto de renda iguais a zero. O resultado de alíquota marginal zero, no modelo original de Mirrlees (1971), aplica-se apenas e tão somente ao indivíduo com a renda mais alta na população. Esse resultado, entretanto, não se aplica ao indivíduo com a segunda renda mais alta e ao restante dos indivíduos, que se deparam com alíquotas marginais positivas.

---

[3] Ver citação acima, p. 3-4.

Nesse sentido, a alíquota marginal zero no topo da distribuição de renda atua como um limite superior na determinação das alíquotas marginais ótimas quando a distribuição de capacidade produtiva tem um limite superior. Como observado por Cremer (2017), "the zero marginal tax rate at the top goes away (even asymptotically) when the ability distribution is unbounded".

*Imposto sobre a renda com alíquota linear*

Da mesma forma, a utilização de um imposto de renda linear não implica, necessariamente, ausência de progressividade na tributação da renda. Essa visão equivocada, de que um imposto de renda com alíquota marginal única resulta em uma tributação passiva relativamente à equidade, ignora que, em geral, esquemas de tributação linear da renda incorporam a realização de transferências, do tipo *lump-sum* ou não, que assumem o papel redistributivo. Um exemplo clássico é aquele de um imposto de renda negativo, que combina um imposto de renda de alíquota marginal única e uma transferência *lump-sum* uniforme paga a todos os contribuintes. Esse esquema é redistributivo, pois, enquanto a alíquota marginal de imposto permanece constante, a alíquota média é negativa para famílias de baixa renda em razão do pagamento da transferência.

*Não tributação do capital*

Finalmente, em relação à tributação do capital, na teoria da tributação ótima não há prescrição incondicional de tributação zero para a renda do capital. O que a teoria enfatiza, corretamente, é que poupança e investimento são decisões de consumo futuro. Portanto, tributá-los equivale a tratar o mesmo fenômeno, consumo, diferenciadamente ao longo do ciclo de vida. Isso gera distorções na decisão entre consumo presente e consumo futuro (poupança/investimento), que podem resultar em ineficiências e perdas de bem-estar.

Esse argumento, entretanto, não se traduz em uma prescrição geral de tributação zero da renda do capital. O resultado de tributação zero da renda do capital é obtido apenas para o caso em que há separação fraca entre trabalho e consumo, preferências homogêneas e com consumo presente e consumo futuro igualmente complementares a lazer (não trabalho), que é o resultado encontrado no Teorema de Atkinson-Stiglitz (1976), ou, em um contexto de decisão intertemporal em que os agentes têm períodos infinitos de vida e a alíquota do imposto tende assintoticamente a zero (Judd, 1985; Chamley, 1986). Esse tipo de resultado é considerado como um *benchmark*, em relação ao qual pode-se investigar a desejabilidade de tributação positiva da renda do capital quando as hipóteses originais são modificadas.

## 4. Comentários finais

Em suma, não há nada na literatura de tributação ótima, tanto nos trabalhos originais quanto em desenvolvimentos posteriores, que advogue, por princípio, a exclusão de objetivos redistributivos no sistema tributário. A existência do inevitável *trade-off* entre equidade e eficiência não sugere essa exclusão. Pelo contrário, o arcabouço da teoria de tributação ótima foi desenvolvida de forma a poder incorporar considerações distributivas na análise tributária e permitir um balanceamento entre os critérios de equidade e eficiência. Frequentemente esquece-se que, na teoria de tributação ótima, a justificativa de se usar tributos distorcivos é justamente a hipótese da existência de uma preocupação com redistribuição. Caso essa preocupação não existisse, o governo poderia fazer uso de um tributo *lump-sum* para gerar a receita necessária sem resultar em distorções e custos de eficiência.